\documentclass{article}

\usepackage{authblk}
\usepackage{geometry}
\usepackage{amsmath,amssymb,amsthm}
\usepackage{graphicx}
\usepackage{algorithm2e}
\usepackage{hyperref}
\newtheorem{theorem}{Theorem}
\newtheorem{definition}{Definition}
\newtheorem{observation}{Observation}
\newtheorem{remark}{Remark}
\title{Computing the Delaunay Graph in High Dimensions}
\author[1]{Tyler H.\ Chang}
\affil[1]{Argonne National Laboratory, 9700 S.\ Cass Ave., Lemont, IL 60439, USA\\
\url{http://thchang.github.io}\\\texttt{thchang@vt.edu}\\
\url{https://orcid.org/0000-0001-9541-7041}}
\date{}

\begin{document}

\maketitle

\begin{abstract}
The Delaunay graph is a $d$-dimensional spatial network, induced by
connecting pairs of spatially-oriented nodes that share an edge in their
Delaunay triangulation.
Current state-of-the-art software for computing the Delaunay graph in
arbitrary dimensions requires computation of the entire $d$-dimensional
Delaunay triangulation, which is infeasible in even moderate dimensions.
Therefore, recent works have proposed individually checking each potential
connection in the graph when the problem dimension is sufficiently high.
This edge-checking strategy is appropriate in settings where the problem
dimension is moderate or high, but the number of nodes is relatively low.
This paper presents the first thorough algorithmic analysis of the
edge-checking strategy, which leads us to recommend a new value for its
look-ahead batch size parameter.
We are also able to show that properly setting this batch size parameter
drastically improves real-world performance, although the exact optimal
value for this parameter is surprisingly difficult to determine.
We provide an open-source implementation of our modified edge-checking
algorithm, and empirically compare this against existing techniques for
calculating the Delaunay graph, showing when this strategy becomes more
efficient than the alternatives.
Next, we demonstrate that our proposed algorithm achieves strong
parallel scaling on a real-world dataset.
Finally, based on anomalies in our performance results, we use our
edge-checking code to investigate the topology of high-dimensional
Delaunay graphs, revealing an unusual pattern in the number of
simplices needed to cover the graph.
\end{abstract}


\noindent
\textbf{Supplement:} \url{https://github.com/thchang/delaunay-graph}

\noindent
\textbf{Funding:} This material is based upon work supported by the
U.S.\ Department of Energy, Office of Science,
Office of Advanced Scientific Computing Research,
SciDAC program under contract number DE-AC02-06CH11357.

\noindent
\textbf{Acknowledgements:} The authors gratefully acknowledge the computing
resources provided on Bebop, a high-performance computing system
operated by the Laboratory Computing Resource Center at
Argonne National Laboratory.

\section{The Delaunay Graph and its Computation}
\label{sec:intro}

Consider a set of $n$ spatial nodes ${\cal P}$ in $\mathbb{R}^d$.
Then the Delaunay triangulation of ${\cal P}$, denoted DT$({\cal P})$,
is a simplicial mesh of the convex hull of ${\cal P}$,
with vertices in ${\cal P}$,
such that every simplex in DT$({\cal P})$ satisfies the
{\sl open circumball property}.

\begin{definition}\label{def:open-circumball}
Let ${\cal S}$ be a $d$-simplex with vertices in ${\cal P}$,
and let ${\cal B}_{\cal S}$ denote the minimum-radius open
$d$-ball whose boundary intersects the vertex set of ${\cal S}$.
Then ${\cal S}$ satisfies the open circumball property if
${\cal B}_{\cal S} \cap {\cal P} = \emptyset$.
\end{definition}

\begin{definition}\label{def:delaunay-triangulation}
The triangulation T$({\cal P})$ is a Delaunay triangulation
if every simplex ${\cal S} \in {\text T}({\cal P})$ satisfies
the open circumball property.
\end{definition}

In this paper, we are interested in constructing the
Delaunay graph DG$({\cal P})$,
defined by connecting all nodes that share an edge
in the DT$({\cal P})$.

\begin{definition}\label{def:delaunay-graph}
The Delaunay graph DG$({\cal P})$ is a spatial network with
nodes in ${\cal P}$, which contains the edge $(x^{(1)},x^{(2)})$
for $x^{(1)},x^{(2)}\in{\cal P}$, if and only if $x^{(1)}$ and
$x^{(2)}$ share an edge in DT$({\cal P})$.
\end{definition}

The applications of DG$({\cal P})$ are numerous, but one motivating
application is to impose topological structures on discrete point
sets in high dimensions.
For example, the Delaunay graph has been explicitly proposed
for many-objective Pareto front analysis \cite{deshpande2016}.

\paragraph*{Important properties}

We now summarize several important properties of
DT$({\cal P})$, which will be used in our analysis.
For further reading on properties of the Delaunay triangulation
in $\mathbb{R}^d$, we recommend \cite[Chapter~4]{cheng2012}.

First, note that so long as ${\cal P}$ do not lie in an affine subspace of
dimension less than $d$, then DT$({\cal P})$ exists.
Furthermore, when the nodes ${\cal P}$ are in {\sl general position}, then
DT$({\cal P})$ -- and therefore also DG$({\cal P})$ -- is unique.

\begin{definition}\label{def:general-position}
Given a set of $n$ nodes ${\cal P}$ in $\mathbb{R}^d$,
we say that ${\cal P}$ are in general position if
(1) DT$({\cal P})$ exists and (2) there
are no sets of $d+2$ or more nodes in ${\cal P}$ that
are cospherical.
\end{definition}

If ${\cal P}$ are drawn from any distribution whose support
in $\mathbb{R}^d$ has nonzero Lebesgue measure, then ${\cal P}$
are in general position with probability one.
However, for many real-world datasets, such as nodes on a uniform grid,
this assumption may not hold.

Second, if ${\cal P}$ are in general position, then any
simplex that satisfies the open circumball property is also a member
of DT$({\cal P})$.

\begin{observation}\label{obs:delaunay-simp}
Assume that ${\cal P}$ are in general position.
For any $d$-simplex ${\cal S}$, if
${\cal B}_{\cal S} \cap {\cal P} = \emptyset$
then ${\cal S}\in{\text DT}({\cal P})$.
\end{observation}

This observation follows immediately from the duality between
the Delaunay triangulation and Voronoi diagram, which is
described in \cite{aurenhammer2013}.

\paragraph*{Complexity of the Delaunay triangulation}

In the worst case, the total number of simplices in DT$({\cal P})$,
could be ${\cal O}(n^{\lceil \frac{d}{2} \rceil})$ \cite{klee1980}.
However, in the average case, the size of DT$({\cal P})$ tends to
grow only linearly in $n$, although it still grows
exponentially in $d$ \cite{dwyer1989}.
This explosion in size is associated with the
{\sl curse of dimensionality} and makes the complete Delaunay
triangulation practically incomputable for $d$ greater
than $\sim 8$ \cite{boissonnat2009}.

On the other hand, DG$({\cal P})$ is fundamentally limited in
its maximum size, since a graph of $n$ nodes can contain at most
$\left( n \atop 2 \right)$ edges.
This tells us that calculation of DG$({\cal P})$ is a
fundamentally easier problem than computation of DT$({\cal P})$.

\paragraph*{Existing Algorithms}

The only known software that explicitly computes and stores
DG$({\cal P})$ in arbitrary dimensions, is the $d$-dimensional
triangulation package \cite{boissonnat2000}
in the Computational Geometry Algorithms Library (CGAL) \cite{fabri2009}.
In this package, the complete Delaunay triangulation is computed via
incremental insertion of nodes, while storing only a small number of
``cached'' simplices, along with DG$({\cal P})$ \cite{boissonnat2009}.

Although it does not explicitly compute DG$({\cal P})$, it is also
worth mentioning the Quickhull algorithm and software \cite{barber1996}.
Quickhull is considered one of the fastest algorithms for computing the
complete Delaunay triangulation in $3$--$5$ dimensions.
From Definition \ref{def:delaunay-graph}, DG$({\cal P})$ can be
constructed from DT$({\cal P})$, as returned by Quickhull.

Both the Delaunay module in CGAL and Quickhull have expected-case time
complexities of ${\cal O}(n \log n + |\text{DT}({\cal P})|)$, which is
optimal for the problem of computing DT$({\cal P})$.
However, for moderately-sized datasets, it has been shown that Quickhull
cannot scale past $\sim 5$ dimensions due to lack of storage
space \cite{boissonnat2009}.
CGAL fairs slightly better since it does not attempt to hold the entire
triangulation in memory, but its time requirements still grow
exponentially with the dimension \cite{boissonnat2009,cgal2022,hornus2008}.

For applications where the complete Delaunay triangulation is not needed,
it is often possible to compute small numbers of simplices much faster.
For example, the problem of querying a single simplex from DT$({\cal P})$
can be posed as a linear program (LP) \cite{chang2020,chang2018}.
The practical running time depends on the method used to locate a simplex;
the empirical performance of the method used here does not establish a
polynomial-time bound.

In the case where we only want to compute DG$({\cal P})$,
a recent algorithm appeared in the appendix of \cite{chang2022},
showing that this can be achieved in $\left(n\atop 2\right)$ simplex queries,
by checking each possible edge between nodes in ${\cal P}$, to see whether
it is an edge in DT$({\cal P})$.
An equivalent strategy is described informally in the 2004 blog
post \cite{fukuda2004}, where the author describes an equivalent solution
based on geometric programming.
However, one must be careful here since off-the-shelf LP
solvers (as suggested by \cite{fukuda2004}) may not properly handle
degenerate scenarios when ${\cal P}$ are not in general position,
as we will discuss in Section \ref{sec:implementation}, and thus
may not be robust enough for real-world usage.

\paragraph*{Computing the Delaunay graph via edge-checking}

The pseudocode for calculating the DG$({\cal P})$ via edge-checking, as
described in \cite{chang2022}, is given below in
Algorithm~\ref{alg:delaunaygraph}.

\begin{algorithm}
\caption{Compute the Delaunay graph via edge-checking.}
\label{alg:delaunaygraph}
\KwData{${\cal P} = \{x^{(1)}, \ldots, x^{(n)}\}$}
\KwOut{${\bf M} = n\times n$ Boolean-valued connectivity matrix}
{\bf Initialize:} ${\bf M}_{i,j} \gets$ {\bf false} for all $i,j=1$ to $n$\;
\For{$i = 1$ to $n$}
{
    \For{$j = i+1$ to $n$}
    {
        ${\bar x}^{(i,j)} \gets (x^{(i)} + x^{(j)}) / 2$ \;
        $S^{(i,j)}$ $\gets$ the vertex set for a Delaunay simplex containing ${\bar x}^{(i,j)}$ \;
        \If{$x^{(i)} \in S^{(i,j)}$ {\bf and} $x^{(j)} \in S^{(i,j)}$}
        {
            ${\bf M}_{i,j}, {\bf M}_{j,i} \gets$ {\bf true} \;
        }
    }
}
\end{algorithm}

\begin{theorem}[Chang et al.\ \cite{chang2022}]
\label{thm:delaunaygraph}
{If ${\cal P}$ is in general position, then Algorithm \ref{alg:delaunaygraph}
computes the connectivity matrix for DG$({\cal P})$.}
\end{theorem}

The proof of Theorem~\ref{thm:delaunaygraph} is in the Appendix.

\begin{remark}
\label{rmk:uniqueness}
If $DG({\cal P})$ is not unique, then we must amend the statement above.
Each edge that is returned by Algorithm~\ref{alg:delaunaygraph} is contained
in {\sl a} Delaunay triangulation;
and otherwise, there exists {\sl a} Delaunay triangulation that does not
contain this edge.
However, Algorithm~\ref{alg:delaunaygraph} may produce a connectivity matrix
${\bf M}$ that is not consistent with any single Delaunay triangulation.
Still, certain useful subgraphs of DG$({\cal P})$
(e.g., the {\sl Gabriel graph}) are subgraphs for {\sl every} possible
Delaunay triangulation, and
these graphs will always be embedded in the connectivity structure computed by
Algorithm~\ref{alg:delaunaygraph}.
\end{remark}

\begin{remark}
\label{rmk:existence}
If DT$({\cal P}$) does not exist, then ${\cal P}$ is contained
in a $(d-1)$-or-lower-dimensional affine subspace.
In \cite{chang2022}, the authors recommend handling this situation by
performing dimension reduction via principle component analysis (PCA) and
calculating DG$({\cal P})$ in the reduced subspace.
\end{remark}

Finally, note that DG$({\cal P})$ is the $1$-skeleton of DT$({\cal P})$.
Algorithm \ref{alg:delaunaygraph} could be generalized to compute any
$\kappa$-skeleton, by instead checking the Barycenter of all
$\sum_{i=1}^\kappa\left(n \atop i\right)$ possible faces.
However, both the time and storage requirements would grow
exponentially in $\kappa$.
Therefore, in this paper, we will be focused exclusively on the $1$-skeleton
DG$({\cal P})$.

\paragraph*{Organization and our contributions}

The remainder of this paper and our novel contributions to this work are
as follows.
In Section \ref{sec:analysis}, we fully {\bf analyze} the effective
computational complexity of Algorithm \ref{alg:delaunaygraph} under a
variety of implementation options.
Next in Section \ref{sec:implementation}, we summarize
{\bf our modified implementation} of Algorithm \ref{alg:delaunaygraph}
and perform experiments to determine its optimal look-ahead batch size
parameter, discovering that for certain batch size choice
{\bf the Delaunay graph gets slightly easier to compute in high dimensions}.
In Section \ref{sec:experiments}, we {\bf empirically compare}
our implementation against the $d$-dimensional Delaunay triangulation
module in CGAL, showing the problem sizes where
Algorithm~\ref{alg:delaunaygraph} becomes preferable to computing
the entire triangulation.
In Section \ref{sec:scaling}, we demonstrate the {\bf parallel scaling}
of our implementation on a real-world dataset.
In Section \ref{sec:topology}, we use our code to {\bf probe the topology}
of high-dimensional Delaunay graphs, to investigate computational anomalies
seen in Section~\ref{sec:implementation}.
Finally, in Section \ref{sec:conclusion}, we summarize our findings
and suggest future research directions.

\section{Analysis of the Edge-Checking Algorithm}
\label{sec:analysis}

As acknowledged in \cite{chang2022}, the primary cost of
Algorithm \ref{alg:delaunaygraph} is computing each vertex set $S^{(i,j)}$.
Since ${\cal O}(n^2)$ such vertex sets must be found, the total cost of the
algorithm will be ${\cal O}(n^2 \times t_S)$, where $t_S$ is the time required
to compute such a vertex set.
This problem can be posed as a LP \cite{chang2022,fukuda2004},
where the goal is to identify the {\sl basic solution}.
However, the problem of identifying a basic solution to a LP is considerably
harder than that of solving the LP \cite{megiddo1991}.
In fact, this is equivalent to Smale's 9th ``open problem for
the next century'' \cite{smale1998}.

However, in practice, this problem of finding basic solutions for a LP can
be solved using the simplex method \cite{dantzig1998},
but when ${\cal P}$ is not in general position, then special care must
be taken to ensure that the solution returned by the simplex method is
nondegenerate.
For this reason, in \cite{chang2022}, the DelaunaySparse software
\cite{chang2020} was used for locating each Delaunay simplex.

\subsection{Summary of DelaunaySparse algorithm and its analysis}
\label{sec:delaunaysparse}

In the DelaunaySparse software \cite{chang2020}, a novel approach is taken
for querying the set of Delaunay simplices containing a given set of
interpolation nodes.
While the approach is equivalent to a nonstandard pivoting strategy for
the simplex method, DelaunaySparse takes advantage of the geometric
structure of the problem to
(1) eliminate redundant computations and
(2) gracefully handle degenerate and near-degenerate scenarios.
For each individual interpolation node $q$ queried, the approach is summarized
in Algorithm \ref{alg:delaunaysparse}.

\begin{algorithm}
\caption{Summary of the DelaunaySparse algorithm.}
\label{alg:delaunaysparse}
\KwData{${\cal P} = \{x^{(1)}, \ldots, x^{(n)}\}$,
        $q \in$ ConvexHull$({\cal P})$}
\KwOut{the vertex set of a simplex ${\cal S}^{(k)}$ containing $q$}
{\bf Initialize:} $k \gets 0$ \;
${\cal S}^{(0)} \gets $ Delaunay simplex containing the nearest
point to $q$ in ${\cal P}$ \;
\While{$q\not\in{\cal S}^{(k)}$}
{
    ${\cal F}^{(k)} \gets$ a facet of ${\cal S}^{(k)}$
        from which $q$ is visible \;
    ${\cal S}^{(k+1)} \gets$ Delaunay simplex containing ${\cal F}^{(k)}$
        across from ${\cal S}^{(k)}$ \;
    $k \gets k + 1$ \;
}
\end{algorithm}

Note that Algorithm \ref{alg:delaunaysparse} is essentially a
{\sl visibility walk} from $q$'s nearest neighbor in ${\cal P}$
to $q$ \cite{devillers2001}.
Such a walk is guaranteed to converge in a finite number of flips ($\ell$)
in the case of Delaunay triangulations \cite{edelsbrunner1989}.
DelaunaySparse checks whether a given vertex set defines a simplex satisfying
Definition~\ref{def:open-circumball} via a QR factorization of its
Barycentric interpolation matrix.
Therefore, DelaunaySparse is able to add an additional vertex to any existing
Delaunay facet while maintaining the Delaunay property via a sequence of $n$
rank-1 QR updates.
By Observation~\ref{obs:delaunay-simp}, any $d$-simplex constructed in this
way is Delaunay.

The total cost of $n$ rank-1 QR updates is ${\cal O}(n d^2)$.
Therefore, the total cost of generating the initial simplex via incremental
construction is ${\cal O}(n d^3)$, and the cost of each ``flip''
is ${\cal O}(n d^2)$.
This makes the total cost of Algorithm \ref{alg:delaunaysparse}
${\cal O}(n d^3 + n d^2 \ell)$, where $\ell$ is the length of the walk.
The maximum possible value for $\ell$ is unclear, and no polynomial bound
on the length of this walk is established here.  Previous empirical results
suggest that, when starting from a ``nearby simplex,'' the walk length is
typically insensitive to $n$ \cite{hornus2008} and grows approximately as
$d\log d$ over the tested dimensions \cite{chang2018}.  These observations
do not establish an asymptotic bound.  For $m$ queries, the cost of the
walks and initial simplex constructions is therefore bounded in terms of
their walk lengths by ${\cal O}(m n d^3 + m n d^2\ell)$, where $\ell$
is an upper bound on the length of the walks.  The $d\log d$ behavior is
an empirical model for $\ell$, not a proven guarantee.

Since the cost of Algorithm \ref{alg:delaunaysparse} is clearly nontrivial,
this approach works best when $m << |\text{DT}({\cal P})|$.
Since $|\text{DT}({\cal P})|$ grows exponentially in $d$, the savings
over methods that compute all of DT$({\cal P})$ are most evident
when $d$ is large.
Additionally, since the intermediate simplices ${\cal S}^{(k)}$ are not
needed after completing each flip, the memory required by this approach
is only ${\cal O}(d^2)$ for internal variable storage, plus the storage space
required for the input set ${\cal P}$ and the output of $m$ simplices.


An additional benefit of this approach is that each intermediate simplex
${\cal S}^{(k)}$ constructed by DelaunaySparse is also a Delaunay simplex.
Therefore, when given a batch of $m$ interpolation points, DelaunaySparse
can ``check ahead'' to see if any future interpolation points are contained
in ${\cal S}^{(k)}$.
Since the QR factorization of ${\cal S}^{(k)}$'s barycentric interpolation
matrix is already computed, the additional cost of this look-ahead is only
${\cal O}(m d^2)$ per iteration, which is essentially negligible when
$m < n$.
In numerous real-world applications (including edge-checking),
the interpolation nodes can be clustered or otherwise arranged such that
many future points are found per iteration,
allowing for tremendous savings by using look-ahead \cite{chang2018}.

\subsection{Analysis of edge-checking with DelaunaySparse}
\label{sec:delgraph-analysis}

In the appendix of \cite{chang2022} and in their source code,
the authors apply DelaunaySparse directly to the batch of all
$\left(n \atop 2\right)$ midpoints between nodes in ${\cal P}$.
Na{\"i}vely, following the analysis from \cite{chang2020}, this results
in an overall time estimate of ${\cal O}(n^3 d^2(d+\ell))$ and
space complexity of just ${\cal O}(n^2)$.
This is derived by simply substituting $m \approx n^2$ into the analysis from
Section \ref{sec:delaunaysparse}.

However, this misses that when $m \in \Theta(n^2)$, then
the complexity of the look-ahead becomes nonnegligible.
In fact, in each iteration $k$, DelaunaySparse will need to look-ahead at
${\cal O}(n^2)$ simplices with complexity ${\cal O}(n^2 d^2)$.
Accounting for both initial simplex construction and look-ahead gives an
overall upper bound of ${\cal O}(n^3 d^3 + n^4 d^2\ell)$ for
Algorithm \ref{alg:delaunaygraph}, where $\ell$ bounds the walk lengths.
The look-ahead term is a worst-case bound; finding multiple midpoints in a
simplex can reduce the work in practice.

The obvious solution to this issue would be to individually pass each of the
midpoints ${\bar x}^{(i,j)}$ from Algorithm \ref{alg:delaunaygraph} to
DelaunaySparse, which would bring the overall time complexity back down to
${\cal O}(n^3 d^2(d+\ell))$.
However, for the edge-checking problem, each simplex constructed contains
many future edges from DG$({\cal P})$, and {\sl some} look-ahead can be an
extremely valuable and time-saving code optimization.
Thus, we instead note that any batch size where we pass no more than
${\cal O}(n)$ midpoints per call to DelaunaySparse does not increase the
overall big-Oh time complexity of Algorithm \ref{alg:delaunaygraph}.

Thus, we hypothesize that the optimal look-ahead batch size is
${\cal O}(n)$, and our first empirical study in Section \ref{sec:batchsize}
will be to test various look-ahead batch sizes in the class ${\cal O}(n)$.

\section{A Modified Implementation}
\label{sec:implementation}

For our implementation, we have forked the peer-reviewed
{\tt Fortran 2008} software VTMOP at \url{https://github.com/vtopt/VTMOP},
from \cite{chang2022}.
The edge-checking Delaunay graph code is located in an independent subroutine
{\tt DELAUNAYGRAPH}, which calls the {\tt DELAUNAYSPARSES} (serial) or
{\tt DELAUNAYSPARSEP} (parallel) subroutines from
DelaunaySparse \cite{chang2020},
located at \url{https://github.com/vtopt/DelaunaySparse}.
In order to utilize these packages without duplicating code,
we have linked both as {\sl submodules} of our GitHub project.

In our repository, we have extracted only the necessary subroutines and
interfaces from VTMOP, which are needed to define the {\tt DELAUNAYGRAPH}
subroutine.
Then, we have made the following modifications:
\begin{itemize}
\item In the original code, DelaunaySparse was applied to all
$\left(n \atop 2\right)$ midpoints in a single batch.
We have broken this call up into multiple batches, each of size
{\tt IBATCH}, where {\tt IBATCH} is a user-provided hyperparameter.
\item In the original code, {\tt DELAUNAYSPARSEP} is used
when {\tt DELAUNAYGRAPH} is called in parallel execution mode.
Due to the blocking of threads during look-ahead, it has been shown that
{\tt DELAUNAYSPARSEP} may not achieve perfect strong scaling for certain
problem sizes \cite{chang2020}.
Since we have already broken the problem up into many small independent
batches, we solve each batch with an independent call
to the serial subroutine {\tt DELAUNAYSPARSES},
using OpenMP with dynamic scheduling to parallelize these calls during
parallel execution mode.
This prevents any possibility of blocking between threads.
\end{itemize}

\paragraph*{A Note on Degeneracy}

In computational geometry software, great care must be taken
to handle degenerate and near-degenerate scenarios (i.e,. when ${\cal P}$
is within an $\varepsilon$-magnitude perturbation of degeneracy).
In the case of CGAL, these scenarios are largely avoided since
CGAL's $d$-dimensional Delaunay module layers over an exact arithmetic kernel
to prevent numerical roundoff errors from perturbing the
solutions \cite{boissonnat2009}.
In many other packages, including Quickhull \cite{barber1996}, some variation
of random or deterministic perturbation is intentionally applied to
ensure nondegeneracy up to the machine precision \cite{edelsbrunner1990},
while maintaining backward stability.

In the case of DelaunaySparse (and by association VTMOP), conditions
that could lead to degeneracy (up to a user-defined tolerance for
$\varepsilon$)
are carefully checked for and handled on a case-by-case basis \cite{chang2020}.
Effectively, DelaunaySparse provides a backward stability guarantee in
the sense that for each vertex set returned by
DelaunaySparse, this set corresponds to a true Delaunay simplex from the
triangulation DT$({\cal \hat P})$, where for each
${\hat x}^{(i)} \in {\cal \hat P}$ corresponding to $x^{(i)} \in {\cal P}$,
$\|{\hat x}^{(i)} - x^{(i)}\|_2 < \varepsilon$.
Additionally, in the {\tt DELAUNAYGRAPH} subroutine from VTMOP,
non-existence is automatically handled via PCA, as described in
Remark \ref{rmk:existence}.

As discussed in Remark \ref{rmk:uniqueness},
the guarantees of the edge-checking code are considerably weaker than
those of DelaunaySparse.
While every simplex from which we calculate our solution is backward
stable in the sense that DelaunaySparse is backward stable, it is impossible
to guarantee that all simplices come from the same Delaunay triangulation
when uniqueness does not hold (up to $\varepsilon$ perturbations).
Therefore, it is possible that the solution may not be consistent with
any particular instance of DG$({\cal P})$, which could be a significant
drawback in certain contexts.

\subsection{Determining the optimal batch size}
\label{sec:batchsize}

As mentioned in Section~\ref{sec:delgraph-analysis}, we expect the
look-ahead batch size parameter {\tt IBATCH} to have a significant impact
on practical performance, with an optimal batch size in ${\cal O}(n)$.
In this section, we experiment with different batch sizes
to identify a value for {\tt IBATCH} that achieves good uniform performance.

All experiments in this section were performed on the Broadwell
nodes of the high-performance
computing (HPC) system Bebop, operated by the Laboratory Computing
Resource Center at Argonne National Laboratory.
We built our code using Intel's {\tt ifort} (v.\ 17.0.4) compiler
with the {\tt -O3} compiler optimization flag and using Intel's Math Kernel
Library (MKL) for all linear algebra computations.

In our initial batch size study, we use four problem sizes,
generated by all pairwise combinations of two possible values for $d$
($4$ and $16$) and $n$ ($200$ and $1000$).
%
Since we expect an optimal look-ahead batch size in ${\cal O}(n)$,
we experiment with batch sizes of the form $C n$, where $C$ is a constant.
For each problem size, we tested values of $C$ ranging from $\frac{1}{4}$
up to $4$, by increments of $\frac{1}{4}$.

\begin{figure}[h]
\includegraphics[width=0.49\textwidth]{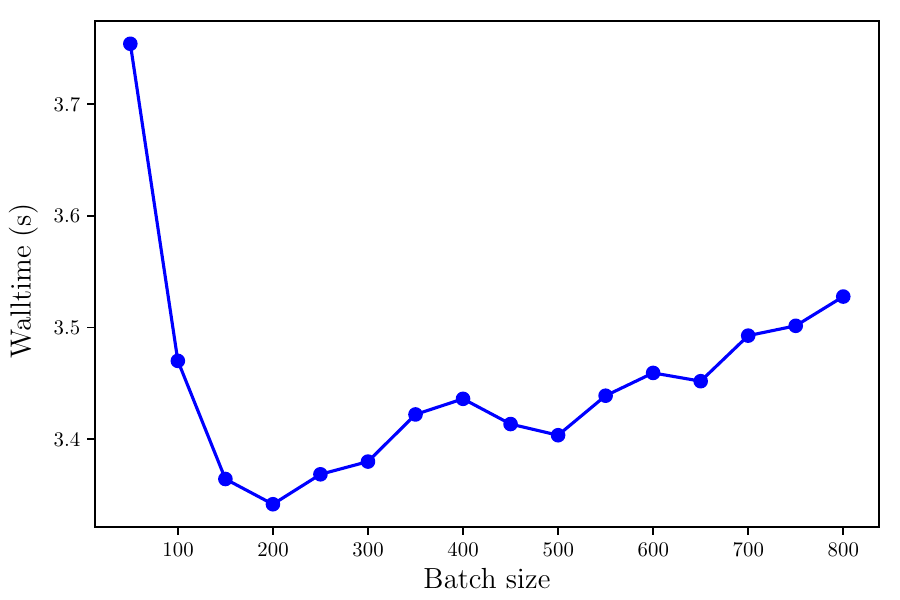}
\includegraphics[width=0.49\textwidth]{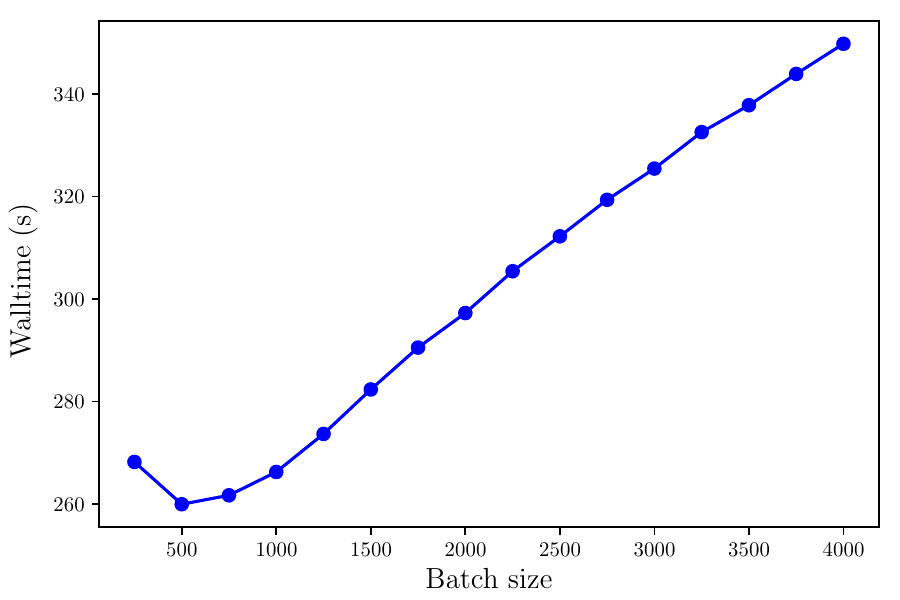}\\
\includegraphics[width=0.49\textwidth]{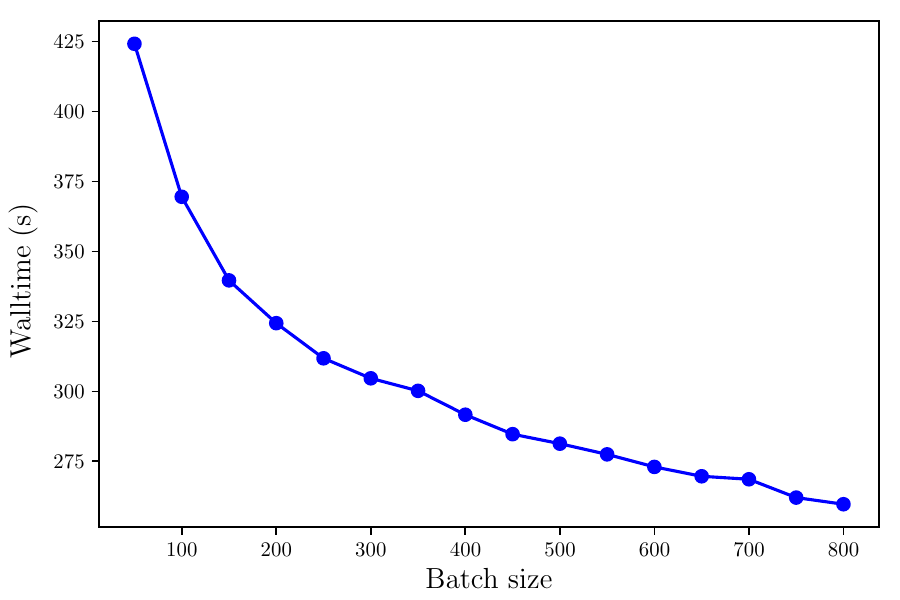}
\includegraphics[width=0.49\textwidth]{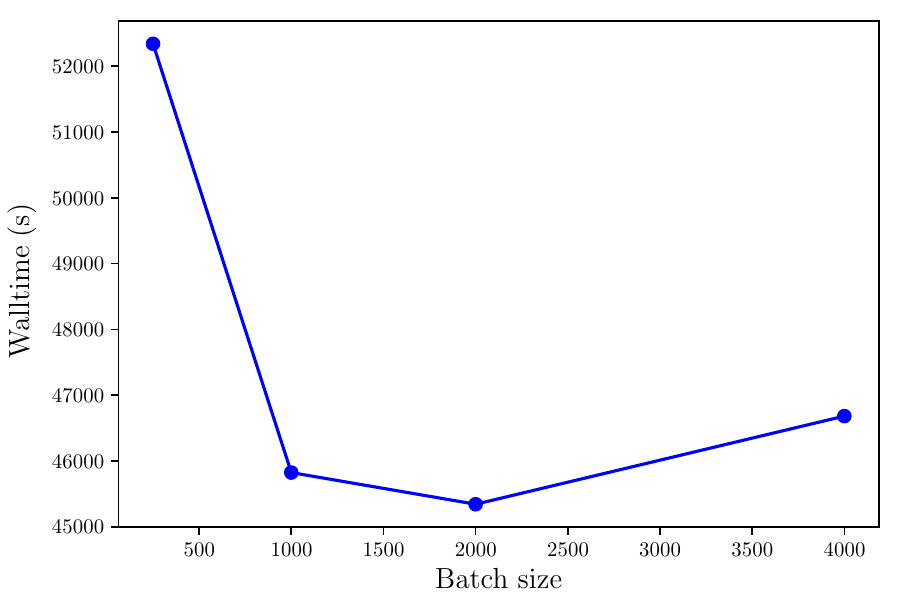}
\caption{Batch sizes used vs.\ average walltimes (in seconds)
for computing DG$({\cal P})$ on various dataset dimensions ($n$ and $d$).
Starting from the top left and moving clockwise, the data sizes are
$d=4$, $n=200$; $d=4$, $n=1000$; $d=16$, $n=1000$; and $d=16$, $n=200$.}
\label{fig:batchsize}
\end{figure}

We tested on data sets that were sampled pseudo-randomly from a uniform
distribution in $[0, 1]^d$ and average our performance over 5 repeated
samples.
Because the largest problem size ($d=16$, $n=1000$) takes a significant
amount of time to solve, we performed the experiments on a reduced
number of batch sizes for this problem, still spanning the same range
and with 5 trials each.
The resulting walltimes, are shown in Figure \ref{fig:batchsize}.

We notice in Figure \ref{fig:batchsize} that, as expected, the batch size
has a significant impact on the performance of the edge-checking algorithm.
Although there is some variation in the exact optimal
value of $C$, $C = 1$ is good for {\sl almost} all problems.
However, for the problem size $(d=16, n=200)$, it appears that no value
of $C\in[\frac{1}{4}, 4]$ is optimal.
This is a peculiar observation, prompting further study.

To further probe the optimal look-ahead batch size, we have performed
the same 5 repeated trials listed above with $d = 4, 8, 12, 16$, and $20$
and $n= 100, 200, 300$, and $400$.
Since the anomaly occurred with a small value of $n$ and the problem is
too expensive to repeat many batch sizes for large values of $n$,
we have stuck with values of $n \leq 400$.
Also, in order to cover a wider range of batch sizes, we use an
exponential spacing of $8$ batch sizes of the form $Cn$, with
$C = \frac{1}{4}, \frac{1}{2}, 1, 2, 4, 8, 16, 32$, plus a $9$th batch size
of $\left(n \atop 2\right)$ -- i.e., no batching.
The optimal batch sizes and their corresponding walltimes are presented in
Table~\ref{tab:batchsize}.

\begin{table}[h]
\caption{Optimal look-ahead batch sizes (based on grid search)
and corresponding average walltimes
(in seconds) to compute DG$({\cal P})$ at varying problem sizes.
Results are given in the format: optimal batch size/average walltime.}
\begin{center}
\begin{tabular}{c|ccccc}
    &           &      $n$   &            &            \\
$d$ &       100 &        200 &        300 &        400 \\
\hline
4 &        0.5 /     3200 &        3.5 /      200 &
           9.8 /      300 &       21.3 /      400 \\
8 &        2.0 /      100 &       15.4 /      100 &
           48.1 /       75 &      106.4 /      100 \\
12 &       11.1 /     4950 &      168.4 /     3200 &
           681.1 /     1200 &     1468.2 /      800 \\
16 &       10.5 /     4950 &      171.2 /    19900 &
           809.2 /     9600 &     2370.3 /     3200 \\
20 &       11.5 /     4950 &      197.6 /    19900 &
           998.5 /    44850 &     2921.6 /    12800 \\
\end{tabular}
\end{center}
\label{tab:batchsize}
\end{table}

The trend in Table~\ref{tab:batchsize} is somewhat unusual.
Although we have not investigated large values of $n$, it appears that
independently of $n$, the
optimal batch size is ${\cal O}(n)$ when $d=4$, $\sim 100$ when $d=8$,
then begins increasingly exponentially for $d>8$ --
eventually reaching the maximum of $\left(n \atop 2\right)$.
Interestingly, this batch size of $\sim 100$ when $d=8$ is independent of
$n$, although the base rate of the exponential increase when $d>8$ appears
smaller for large values of $n$.
It is unclear to what extent these values depend on machine-specific
constants such as cache sizes.
However, this trend is fundamentally interesting, so we present additional
empirical findings on how it relates to the topology
of DG$({\cal P})$ in Section~\ref{sec:topology}.

A batch size of $n$ agrees with our analysis from
Section~\ref{sec:delgraph-analysis} and causes the empirical performance
to match our analysis, although it is far from optimal for large values of
$d$ (see Appendix).
Since we were not able to find a meaningful function to predict the
optimal batch size,
we set the default value for the {\tt IBATCH} parameter to
$n$ and use this value for all future experiments.
We acknowledge that significant performance improvements could be obtained
by choosing an exact optimal batch size for each problem dimensions.

\subsection{Comparison against VTMOP}
\label{sec:dg-vtmop}

In order to verify that our modifications result in real improvements
at a variety of problem sizes, we present a brief comparison between our
modified edge-checking code with {\tt IBATCH}=$n$ (labeled DelaunayGraph
throughout) and the original {\tt DELAUNAYGRAPH} subroutine in
VTMOP \cite{chang2022} (labeled VTMOP throughout).

All experiments in this section are carried out on the same Broadwell nodes
of the HPC system Bebop described in Section \ref{sec:batchsize}, and
both of the codes have been built using the same version of the
{\tt ifort} compiler, with the same linear algebra libraries and
compiler options.
Table~\ref{tab:dg-vtmop} displays the recorded walltime in seconds at
a variety of dimensions and problem sizes, averaged over 5 data sets
uniformly sampled from $[0,1]^d$, just as before.
Entries whose walltime-to-solution exceeds 86,400 secs (24 hrs) were not
completed due to time limitations on Bebop.

\begin{table}[h]
\caption{Average walltime in seconds to compute DG$({\cal P})$ with
DelaunayGraph vs.\ VTMOP, for uniformly distributed nodes in $[0,1]^d$.
Results are given in the format: DelaunayGraph time / VTMOP time.
Entries that read ``time'' exceeded our compute budget of 24 hours.
Entries where DelaunayGraph outperforms VTMOP are highlighted in bold.}
\begin{center}
\begin{tabular}{c|ccccc}
    &           &            &       $n$  &            &            \\
$d$ &       200 &        600 &      1,000 &      1,400 &      1,800 \\
\hline
 4  &  {\bf 3.5 /        4.5}&{\bf   65.2 /      131  }&
     {\bf 268   /      619  }&{\bf  695   /    1,721  }&
   {\bf 1,416   /    3,671  }\\
 8  & {\bf 16.2 /      146  }&{\bf  411   /   21,232  }&
   {\bf 1,835   /      time} &{\bf 4,917   /     time} &
  {\bf 10,327   /      time} \\ 
 12 &     201   /      201   &{\bf 5,182   /  36,418  }&
  {\bf 22,424   /       time}&{\bf 56,277   /     time}&
           time /       time \\ 
 16 &     303   /      183   &{\bf 8,876   /  22,866  }&
  {\bf 42,366   /       time}&       time /       time &
           time /       time \\
 20 &     436   /      176   &{\bf 13,618   / 20,584  }&
  {\bf 66,754   /       time}&       time /       time &
           time /       time \\
\end{tabular}
\end{center}
\label{tab:dg-vtmop}
\end{table}

In Table~\ref{tab:dg-vtmop}, our modified DelaunayGraph code
outperforms the original VTMOP code at {\sl almost} all problem sizes.
As expected, the savings are greatest for large values of $n$
consistent with the reduction in the look-ahead work described in
Section~\ref{sec:delgraph-analysis}.
However, VTMOP can slightly outperform our code for larger values of $d$
and smaller values of $n$, due to the optimality of large batch sizes in
these scenarios.

Additionally, in the case of VTMOP, something surprising happens in the
larger dimensions for the smallest problem size.
When $n=200$ and $600$ (and possibly for larger problem sizes as well),
the computation time stops growing and even {\sl decreases} for $d>12$.
One possible explanation is that more future midpoints are discovered
during the simplex walk at these problem sizes.  If the empirical model
$\ell \approx d\log d$ applies, the cost per walk also grows with $d$;
the observed decrease in runtime reflects the combined effect of these
costs and the number of walks.  These finite-size observations do not
establish an asymptotic rate.
This trend hints that DG$({\cal P})$ has an interesting topology for large
values of $d$, which we will explore in Section~\ref{sec:topology}.

\section{An Empirical Comparison of Delaunay Graph Software}
\label{sec:experiments}

In this section, we compare our modified edge-checking software
(referred to throughout this section as DelaunayGraph) against
the $d$-dimensional Delaunay methods in CGAL \cite{boissonnat2009}
(referred to throughout this section as CGAL).
This comparison also takes place on the Broadwell nodes of Bebop, as
described in Section \ref{sec:batchsize}.
We perform our comparisons on the same uniformly sampled datasets as in
Section \ref{sec:dg-vtmop}, and all performance
statistics are based on averages over 5 repeated trials with different
data generations.

Throughout this section, the DelaunayGraph code has been
built using the same version of the {\tt ifort} compiler, linear algebra
libraries, and compiler options as in Section~\ref{sec:implementation}.
We have compared against the CGAL library (v.\ 5.5.1), cloned from
\url{https://github.com/CGAL/cgal} and
built using the GNU {\tt gcc/gpp} compiler (v.\ 8.2.0) with
Boost v.\ 1.76.0,
GNU multiple precision (GMP) arithmetic library v.\ 6.1.2,
GNU multiple precision floating-point reliable (MPFR) library v.\ 4.0.2,
and Eigen v.\ 3.3.9.

\subsection{Results of serial comparison against CGAL}

Table~\ref{tab:dg-cgal} compares the average walltime in seconds between
our modified DelaunayGraph code and CGAL.
Again, walltimes exceeding 86,400 secs (24 hrs) have been omitted
to save compute time.
Additionally, note that although CGAL stores a graph representation
requiring only ${\cal O}(n^2)$ space plus its simplex cache,
it can still exceed the memory limits of
our Broadwell nodes while re-triangulating the conflict region during
vertex insertion \cite{boissonnat2009}.

\begin{table}[h]
\caption{Average walltime in seconds to compute DG$({\cal P})$ with
DelaunayGraph vs.\ CGAL, for uniformly distributed nodes in $[0,1]^d$.
Results are given in the format: DelaunayGraph time / CGAL time.
Entries that read ``swap'' exceeded our memory limitations, and
entries that read ``time'' exceeded our compute budget of 24 hours.
Entries where DelaunayGraph outperforms CGAL are highlighted in bold.}
\begin{center}
\begin{tabular}{c|ccccc}
    &           &            &       $n$  &            &            \\
$d$ &       200 &        600 &      1,000 &      1,400 &      1,800 \\
\hline
 4  &       3.5 /        0.6 &       65.2 /        2.4 &
            268 /        4.4 &      695   /        6.4 &
          1,416 /        8.6 \\
 8  &      16.2 /       16.2 &      411   /      152   &
        1,835   /      377   &    4,917   /      652   &
       10,327   /      977   \\
 12 &{\bf 201   /    9,507  }&{\bf 5,182  /       swap}&
  {\bf 22,424   /       swap}&{\bf 56,277 /       swap}&
           time /       swap \\
 16 &{\bf 303   /       swap}&{\bf 8,876  /       swap}&
  {\bf 42,366   /       swap}&       time /       swap &
           time /       swap \\
 20 &{\bf 436   /       swap}&{\bf 13,618 /       swap}&
  {\bf 66,754   /       swap}&       time /       swap &
           time /       swap \\
\end{tabular}
\end{center}
\label{tab:dg-cgal}
\end{table}

As expected, for small values of $d$, the CGAL code significantly outperforms
the DelaunayGraph code.
However, when $d>8$, we see that the CGAL
code becomes extremely slow for even the smallest problem size,
and runs out of swap-space for all larger problem sizes.
Meanwhile, our DelaunayGraph code is able to continue scaling, although
its runtime becomes large.
We will see in Section~\ref{sec:scaling} that we can continue to scale
the DelaunayGraph code to larger problems by utilizing its parallel
nature, and we can even outperform the CGAL code in terms of walltime
at problem sizes where the serial code is an order of magnitude slower.

\section{Parallel Scaling}
\label{sec:scaling}

The DelaunayGraph implementation from Section \ref{sec:implementation}
is embarrassingly parallel.
Since we perform numerous independent calls to {\tt DELAUNAYSPARSES}
with batch sizes of $n$, we can simply perform multiple calls in parallel,
e.g., using either OpenMP or MPI.
In this section, we demonstrate how the DelaunayGraph implementation from
Section \ref{sec:implementation} can achieve strong scaling on a
single shared-memory node with 64 cores by using OpenMP parallelism.

The runtimes in this section are gathered by running the DelaunayGraph code
on a real-world dataset, described in Section \ref{sec:fayans}.
All runtimes have been recorded on the Knights Landing nodes of the HPC
machine Bebop, described in Section \ref{sec:batchsize}.
It is worth noting that the Knights Landing nodes have significantly worse
serial performance than the Broadwell nodes (used in Sections
\ref{sec:batchsize} and \ref{sec:experiments}),
so we do not expect the serial runtimes to be comparable to those in
Table~\ref{tab:dg-cgal}.
However, the Knights Landing nodes have 64 physical cores, making them
ideal for running the parallel scaling study.
We have also built all of our codes and libraries using the same compilers,
flags, and libraries as described in Sections \ref{sec:batchsize} and
\ref{sec:experiments}.

\subsection{The Fayans EDF dataset}
\label{sec:fayans}

The dataset used in this section consists of the $\chi^2$-error across
9 different observable classes, resulting from over 52,000 unique
parameterizations of the Fayan's energy density functional (EDF) model.
This dataset was collected by \cite{bollapragada2020} for the purpose
of fitting the Fayans EDF model with low error across all 9 observable
classes.

In order to generate the dataset used here, we have rescaled all of
the $\chi^2$-errors into the range $[-1, 1]$ and filtered out a subset of
the $\chi^2$-error values that are
\begin{enumerate}
\item Pareto optimal (for the purpose of minimizing $\chi^2$-values across
all 9 classes),
\item unique up to a tolerance of $10^{-4}$, and
\item do not exceed some predetermined ranges of ``interesting'' values.
\end{enumerate}
Since the $9$-objective Pareto front is actually an $8$-dimensional manifold
in the objective-space, we reduce the dimension of this subset by
applying the projection described in Section 3.1 of \cite{chang2020}.
The resulting dataset, for which we construct DG$({\cal P})$,
consists of just $n=$1,510 nodes in $d=8$ dimensions.
For more information on the significance of this transformed dataset in the
context of multiobjective optimization, see \cite{chang2020,deshpande2016}.

\subsection{Parallel scaling results}

\begin{figure}[h]
\begin{center}
\includegraphics[width=0.85\textwidth]{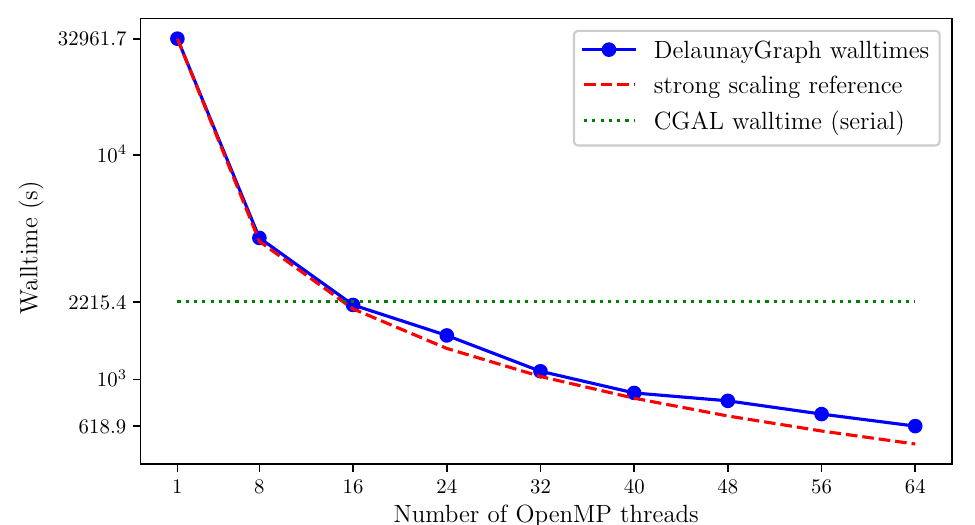}
\end{center}
\caption{
The number of threads used vs.\ the walltime to solution when running the
DelaunayGraph code on the Fayans EDF dataset from \cite{bollapragada2020},
after rescaling objectives and applying the projection from
\cite[Section~3.1]{chang2022}.
For reference, we also show the curve for a perfect strong-scaling and
the serial walltime from solving the same problem with CGAL.
The y-axis is given at a log-scale.}
\label{fig:fayans-scaling}
\end{figure}

The results of running the DelaunayGraph code from
Section \ref{sec:implementation}, using OpenMP parallelism are given in
Figure \ref{fig:fayans-scaling}.
For comparison, we have also solved the same problem serially using CGAL's
$d$-dimensional Delaunay triangulation (which is not parallelizable),
and we provide the perfect strong-scaling curve for reference.

As seen in Figure \ref{fig:fayans-scaling}, we achieve near perfect
strong scaling on this problem.
The CGAL code is initially faster at this problem size on the Knights Landing
nodes with a serial runtime of $2,215.449$ secs (about $37$ mins),
while the DelaunayGraph code requires a serial runtime of $32,961.669$ secs
(just over $9$ hours).
However, with $16$ OpenMP threads, the DelaunayGraph code can approximately
match CGAL's performance, and with all $64$ threads the DelaunayGraph
code can find the solution with a walltime of just $618.9$ secs
(about $10$ mins).

The strong scaling trend indicates that this improvement would continue with
more cores, and since there is no communication between threads, this approach
could even be distributed (e.g., by using MPI).
Given the growth in walltimes shown in Table~\ref{tab:dg-cgal},
a distributed approach
would be necessary for calculating DG$({\cal P})$ for extremely large
datasets in high-dimensions.
Additionally, when $n$ is large, this could allow the ${\cal O}(n^2)$ space
complexity to be fully distributed, although care must be taken in how
the output graph is represented since the full connectivity matrix still has
${\cal O}(n^2)$ size.

\section{Topology of High Dimensional Delaunay Graphs}
\label{sec:topology}

We now revisit the interesting behavior of the VTMOP implementation in
high dimensions, as described in Section~\ref{sec:dg-vtmop}.
The decrease in runtime with increasing $d$ at certain problem sizes
may be related to the number of future midpoints found during each walk.
This is an empirical observation and does not establish a rate of change.

To check this hypothesis and learn how it relates to the overall
topology of DG$({\cal P})$, we have performed the same experiments
presented in Table~\ref{tab:batchsize} using a maximal value for
{\tt IBATCH} (i.e., no batching).
During these experiments, we record both the average number of unique
simplices in the solution returned by DelaunaySparse and the
{\sl edge density ratio}
(i.e., the number of {\tt true} values in the upper triangle of
our connectivity matrix over $\left(n \atop 2\right)$).
The results are presented in Tables~\ref{tab:nsimps} and \ref{tab:nedges}.

\begin{table}[h]
\caption{Average number of simplices computed by DelaunaySparse
when computing DG$({\cal P})$,
for ${\cal P}$ sampled uniformly from $[0,1]^d$.}
\begin{center}
\begin{tabular}{c|ccccc}
    &           &      $n$   &            &            \\
$d$ &       100 &        200 &        300 &        400 \\
\hline
 4 &     1,216.2 &     3,183.2 &     5,379.6 &     7,757.6 \\
 8 &     2,817.0 &    13,560.6 &    33,110.6 &    61,478.8 \\
12 &     1,919.8 &     9,801.2 &    24,949.4 &    47,883.2 \\
16 &     1,405.4 &     7,610.0 &    19,548.0 &    37,690.0 \\
20 &     1,094.6 &     6,171.0 &    16,168.2 &    31,319.0 \\
\end{tabular}
\end{center}
\label{tab:nsimps}
\end{table}

\begin{table}[h]
\caption{Average edge density ratio of DG$({\cal P})$,
for ${\cal P}$ sampled uniformly from $[0,1]^d$.}
\begin{center}
\begin{tabular}{c|ccccc}
    &           &      $n$   &            &            \\
$d$ &       100 &        200 &        300 &        400 \\
\hline
4 &  0.246 &  0.139 &  0.098 &  0.076 \\
8 &  0.801 &  0.634 &  0.532 &  0.463 \\
12 &  0.984 &  0.941 &  0.895 &  0.855 \\
16 &  1.000 &  0.996 &  0.989 &  0.980 \\
20 &  1.000 &  1.000 &  0.999 &  0.999 \\
\end{tabular}
\end{center}
\label{tab:nedges}
\end{table}

From Table~\ref{tab:nsimps}, the number of simplices in the solution
is maximized when $d=8$ then begins to decay.
This confirms our hypothesis and also partially explains why the
optimal batch size takes its minima, independent of $n$, when $d=8$.
However, we still cannot explain why $d=8$ is the maximal dimension
for the number of unique simplices in our solution set.
Since the connectivity of each node increases exponentially with
$d$ \cite{dwyer1989,hornus2008}, we had imagined that
DG$({\cal P})$ might become saturated (reaching 100\% edge density ratio)
when $d=8$, partially explaining the subsequent decay in solution size.
Although it is somewhat surprising how quickly DG$({\cal P})$ becomes
fully saturated with increasing $d$, we do not observe any correlation
between approaching 1.0 density in Table~\ref{tab:nedges} and the maximal
number of simplices observed in Table~\ref{tab:nsimps},
nor is it logical for this latter dimension to be a
constant, independent of $n$.
At this time, we are unable to offer a satisfactory explanation of this
phenomenon and leave it as a topic of future investigation.

\section{Conclusion and Future Work}
\label{sec:conclusion}

In this paper, we have summarized and analyzed an edge-checking algorithm
for computing DG$({\cal P})$. Its running time depends on the length of
the DelaunaySparse simplex walk; the empirical behavior reported for that
walk does not establish a polynomial-time bound.
The edge-checking algorithm is slower than alternatives that rely on the
complete Delaunay triangulation for small $d$, but can greatly outperform
other algorithms for $d>8$.
We provide a Fortran code implementing edge-checking, based
on a modification of \cite{chang2022} and using the DelaunaySparse
package \cite{chang2020}.
The resulting code achieves strong scaling under shared-memory parallelism,
which can help address its high-order time complexity with respect to $n$.

Also note that an earlier version of this paper was submitted to a major
computational geometry conference in 2022/2023 and rejected.  The reviewers
raised several major concerns of which we are grateful.  Several criticisms
that were easily addressable through wording have been addressed in this
revision.  A major criticism that remains unaddressed because of computational
limitations is that much of the experimental scaling analysis and the edge
density ratios are based on uniformly sampled point sets.  Real data
distributions are often nonuniform, and uniform sampling is known to behave
poorly in high dimensions.  We acknowledge that this is a valid limitation,
while maintaining that the fundamental results remain interesting.  A further
scaling study using other data distributions would be worthwhile, but we no
longer have access to the computing resources needed to conduct it.

This paper raises one unanswered question.
It is clear from the results in Sections~\ref{sec:batchsize} and
\ref{sec:dg-vtmop} that a batch size of ${\cal O}(n)$
is not optimal across all values of $d$.
Surprisingly, when a maximal look-ahead depth is used,
the empirical runtime of our code can decrease as $d$ grows.
In Section \ref{sec:topology}, we show that this may be related to
the topology of DG(${\cal P})$ in high dimensions, although
this connection is not fully explained by our observations.
As it may be of fundamental interest in the area of topological data
analysis, we recommend additional study of this phenomenon.



\bibliography{del-refs}

\appendix

\paragraph*{Proof of Theorem \ref{thm:delaunaygraph}}

{\bf Theorem \ref{thm:delaunaygraph}} (Chang et al.\ \cite{chang2022}).
{\sl
If ${\cal P}$ is in general position, then Algorithm \ref{alg:delaunaygraph}
computes the connectivity matrix for DG$({\cal P})$.}

\begin{proof}
Suppose that ${\cal P}$ is in general position, 
let $x^{(i)}, x^{(j)} \in {\cal P}$, and let $S^{(i,j)}$ be the vertex set
of a Delaunay simplex containing their midpoint $\frac{x^{(i)} + x^{(j)}}{2}$,
as in Algorithm \ref{alg:delaunaygraph}.
If $S^{(i,j)}$ contains both $x^{(i)}$ and $x^{(j)}$,
then clearly $x^{(i)}$ and $x^{(j)}$ are connected in DG$({\cal P})$.
Otherwise, since the simplex whose vertices are $S^{(i,j)}$ contains
the midpoint between $x^{(i)}$ and $x^{(j)}$, at least one facet of this
simplex must separate them.
Thus, the simplex with vertices in $S^{(i,j)}$ is a certificate that
$x^{(i)}$ and $x^{(j)}$ are separated in the Delaunay triangulation and
cannot be connected in DG$({\cal P})$.
\end{proof}

\paragraph*{Additional Tables}

\begin{table}[h]
\caption{Look-ahead batch sizes ({\tt IBATCH}) and corresponding average
walltimes (in seconds) to compute DG$({\cal P})$ for varying $d$ when
$n=100$.
Within each column, the fastest time is given in bold.}
\begin{center}
\begin{tabular}{c|ccccc}
                &       &      &   $d$  &       &       \\
{\tt IBATCH}    &     4 &  8   &   12   &   16  &   20  \\
\hline
25 &       0.66 &       2.25 &      29.22 &      45.28 &      63.39 \\
50 &       0.57 &       2.10 &      25.17 &      38.61 &      53.03 \\
100 &       0.53 &  {\bf 2.04}&      22.00 &      32.16 &      43.15 \\
200 &       0.51 &       2.10 &      19.36 &      26.85 &      34.76 \\
400 &       0.51 &       2.26 &      17.15 &      22.50 &      27.93 \\
800 &       0.50 &       2.62 &      15.26 &      18.54 &      22.34 \\
1600 &       0.49 &       3.21 &      13.43 &      15.00 &      17.49 \\
3200 &  {\bf 0.48}&       3.94 &      12.19 &      12.55 &      14.31 \\
4950 &       0.49 &       5.34 & {\bf 11.07}& {\bf 10.52}& {\bf 11.53}\\
\end{tabular}
\end{center}
\label{tab:batchsize-n100}
\end{table}

\begin{table}[h]
\caption{Look-ahead batch sizes ({\tt IBATCH}) and corresponding average
walltimes (in seconds) to compute DG$({\cal P})$ for varying $d$ when
$n=200$.
Within each column, the fastest time is given in bold.}
\begin{center}
\begin{tabular}{c|ccccc}
                &       &      &   $d$  &       &       \\
{\tt IBATCH}    &     4 &  8   &   12   &   16  &   20  \\
\hline
50 &       3.86 &      15.60 &     242.45 &     389.60 &     684.90 \\
100 &       3.65 & {\bf 15.43}&     217.15 &     339.95 &     587.40 \\
200 &  {\bf 3.46}&      16.14 &     198.07 &     298.17 &     505.83 \\
400 &       3.49 &      18.50 &     186.41 &     268.22 &     440.88 \\
800 &       3.62 &      23.11 &     177.09 &     242.03 &     384.84 \\
1600 &       3.79 &      30.89 &     171.21 &     216.90 &     330.91 \\
3200 &       4.02 &      44.26 &{\bf 168.43}&     197.22 &     280.92 \\
6400 &       4.20 &      67.22 &     172.46 &     183.05 &     243.08 \\
19900 &       4.46 &     144.28 &     204.14 &{\bf 171.18}&{\bf 197.59}\\
\end{tabular}
\end{center}
\label{tab:batchsize-n200}
\end{table}

\begin{table}[h]
\caption{Look-ahead batch sizes ({\tt IBATCH}) and corresponding average
walltimes (in seconds) to compute DG$({\cal P})$ for varying $d$ when
$n=300$.
Within each column, the fastest time is given in bold.}
\begin{center}
\begin{tabular}{c|ccccc}
                &       &      &   $d$  &       &       \\
{\tt IBATCH}    &     4 &  8   &   12   &   16  &   20  \\
\hline
75 &      10.76 & {\bf 48.12}&     831.52 &    1314.84 &    2183.60 \\
150 &      10.01 &      49.16 &     757.78 &    1167.11 &    1905.58 \\
300 & {\bf  9.78}&      53.96 &     705.54 &    1050.13 &    1690.32 \\
600 &      10.19 &      65.31 &     685.26 &     974.02 &    1535.22 \\
1200 &      10.99 &      87.23 &{\bf 681.10}&     915.47 &    1390.66 \\
2400 &      12.01 &     127.01 &     692.09 &     861.24 &    1263.49 \\
4800 &      13.33 &     194.36 &     737.22 &     827.11 &    1152.99 \\
9600 &      14.18 &     309.27 &     818.59 &{\bf 809.23}&    1059.61 \\
44850 &      15.58 &     958.99 &    1349.33 &     934.64 &{\bf 998.46}\\
\end{tabular}
\end{center}
\label{tab:batchsize-n300}
\end{table}

\begin{table}[h]
\caption{Look-ahead batch sizes ({\tt IBATCH}) and corresponding average
walltimes (in seconds) to compute DG$({\cal P})$ for varying $d$ when
$n=400$.
Within each column, the fastest time is given in bold.}
\begin{center}
\begin{tabular}{c|ccccc}
                &       &      &   $d$  &       &       \\
{\tt IBATCH}    &     4 &  8   &   12   &   16  &   20  \\
\hline
100 &      22.59 &{\bf 106.45}&    1682.85 &    3304.31 &    4680.78 \\
200 &      21.48 &     111.56 &    1575.92 &    2936.63 &    4088.70 \\
400 & {\bf 21.31}&     125.88 &    1487.25 &    2660.99 &    3659.85 \\
800 &      22.61 &     158.27 &{\bf 1468.18}&   2504.81 &    3376.69 \\
1600 &      24.93 &     221.63 &    1512.44 &    2424.34 &    3211.30 \\
3200 &      28.00 &     337.11 &    1636.16 &{\bf 2370.31}&   3050.96 \\
6400 &      31.32 &     537.58 &    1845.23 &    2403.21 &    2945.79 \\
12800 &      33.83 &     886.81 &    2227.39 &    2465.94 &{\bf 2921.62}\\
79800 &      37.73 &    3548.05 &    5121.66 &    3466.96 &    3372.26 \\
\end{tabular}
\end{center}
\label{tab:batchsize-n400}
\end{table}

\end{document}